"**Cryogenic *In-situ* clamped beam testing of Sn96**"

A. Lupinacci[1], J. Kacher[3], A.A. Shapiro[4], P. Hosemann[5*], A.M. Minor[1,2*]

[1]Department of Materials Science and Engineering, University of California, Berkeley, CA
[2]National Center for Electron Microscopy, Lawrence Berkeley National Laboratory, Berkeley, CA
[3]School of Materials Science and Engineering, Georgia Institute of Technology, Atlanta, GA
[4]Jet Propulsion Laboratory, California Institute of Technology, Pasadena, CA
[5]Department of Nuclear Engineering, University of California, Berkeley, CA
*corresponding authors: A.M. Minor aminor@berkeley.edu,  P. Hosemann peterh@berkeley.edu

**Abstract**

The deformation behavior of materials that exhibit a ductile to brittle transition have traditionally been characterized ex-situ and limited to bulk samples.  Here, we use a unique experimental setup to perform in situ cryogenic mechanical testing of Sn 96 solder alloy in a clamped beam configuration at -142°C, well below the alloy's DBTT.  Electron backscattered diffraction analysis before and after the *in situ* experiments showed how deformation ahead of the crack tip is accommodated by the Sn matrix or intermetallic phase particles.  Similar to behavior observed in pure Sn, within the Sn matrix of the Sn 96 alloy deformation twinning is the primary deformation mechanism below the DBTT.  Similar to the twins previously observed in pure Sn, twinning behavior in the Sn matrix is consistent with the formation of the {301} and {101} twins in Sn.  Within the intermetallic phase in Sn96, slip steps were observed that were not observed prior to bending.

1. Introduction

*In-situ* micromechanical testing has proven to be very helpful in understanding underlying deformation mechanisms that drive global mechanical behavior.  In particular, small scale testing is advantageous for probing materials in an extreme environment, such as elevated or cryogenic temperatures. When coupled with traditional characterization techniques such as electron backscatter diffraction (EBSD) and secondary electron imaging, the deformation mechanisms such as twinning and crack initiation can be detected and directly linked to microstructural characteristics of the material.

Materials behavior at cryogenic temperatures has become increasingly more important for aerospace applications where materials are routinely exposed to these extremes.  In a previous study we focused on the ductile to brittle transition (DBTT) of pure Sn, using *In-situ* micromechanical testing to probe the deformation mechanisms above and below the DBTT[1].  Since Sn is the major constituent of most solders used in industry, our first study focused on the pure Sn with no other alloy element present given that the body centered tetragonal (bct) crystal structure β-Sn dictates the global mechanical properties of most solders[2].  The bct structure gives rise to several unique deformation mechanisms that change as a function of temperature, strain rate and orientation.  At room temperature and relatively low strain rates, Sn deforms by dislocation-mediated slip[3,4].  Below the DBTT, the primary deformation mechanism in pure Sn is deformation twinning. The primary twinning system in β-Sn is the {301}<103] and the {101}<101] systems[5,6], both of which are present and active below the DBTT[1]. A traditional method to characterize the mechanical properties of solder in the low temperature regime is via Charpy impact testing. While Charpy testing is very useful for identifying global trends and establishing the DBTT, it is limited in terms of characterizing the deformation mechanisms that are associated with mechanical behavior as a function of temperature. In-situ micromechanical testing at cryogenic temperature allows us to draw a direct link between the observed deformation and the mechanical response. While our previous work was able to draw a connection between the mechanical behavior of pure Sn above and below the DBTT with its associated deformation mechanisms, the question as to how alloying elements affect this behavior still remains.

In this study Sn96.5Ag3.5 was investigated in order to better understand the effect that alloying has on the DBTT. Known as 'Sn96', this alloy is considered a high temperature solder as the eutectic melting temperature is at 221°C. Unlike pure Sn which has a DBTT of -125 °C[1], Sn96 exhibits a DBTT at a warmer temperature near -45°C[7]. Due to the dispersed $Ag_3Sn$ in the Sn matrix, SnAg solder alloys exhibit excellent mechanical properties and high reliability.[8]

Sn96 is also commonly used within the aerospace community, making the DBTT relevant to microelectronic reliability where this alloy is used. Solder joints are used extensively in packaging of electronics. These joints provide both electrical connections and mechanical support for the electrical package. Differences in the thermal expansion coefficients of materials used in electronic packaging cause cyclical strains in the solder joints upon thermal cycling[1-2]. Reliability of the solder joint depends upon a number of mechanical and microstructural evolutionary processes that occur interactively during its service lifetime. This underscores the importance of understanding the deformation mechanisms of these materials where material must both demonstrate both mechanical integrity and reliability at extreme temperatures. Furthermore, while SnPb eutectic solder is still one of the most widely used solder alloys, due to the hazards that Pb poses to the environment, many members of the microelectronic community have started gravitating away from Pb based solders[9]. Recent government regulations have imposed restrictions on the use of lead, the European RoHS (Reduction of Hazardous Substances) banned the use of lead in July 2006 and the US Environmental Protection Agency has also expressed concerns on the usage of lead due to its inherent toxic behavior. Therefore, making it even more important that we fully characterize and understand the behavior of Pb-free solders at low temperature. Since the functionality and lifespan of an electronic product is directly linked to the durability and reliability of its solder joints, it is necessary to fully understand and characterize the underlying deformation mechanisms of the solder material[10].

The objective of this study is to better understand the effect that alloying elements have on the deformation mechanisms responsible for the ductile to brittle transition in solder alloys such as Sn 96. Of particular interest is the effect that intermetallic particles and grain boundaries have on behavior of an existing defect or in this case the crack tip of a clamped beam specimen. Instead of using a pillar geometry as demonstrated in our previous study[1], this study uses a suitably loaded edge notched doubly clamped beam as demonstrated by Jaya et al.[11] Fracture tests are primarily characterized by the competition between the available driving force (G) and the resistance (R) offered by the material to experience crack growth. The possible geometries and experimental constraints that can be used to achieve stable cracking in brittle materials has been extensively discussed by Mai and Atkins[12]. In order to observe the crack trajectory, there needs to be inherent geometric stability in the sample configuration, especially when the specimen being tested is brittle[13]. The edge-notched clamped beam bend test geometry has been evaluated and shown stability even under load control.[13-16.] This test geometry has been modified from a traditional three point bend test to accommodate the constraints imposed in micro-scale testing.[16.] Furthermore, the constraints this geometry introduces are more practical and of interest in evaluating deformation mechanisms that could be relevant in solder joints which are traditionally constrained in a microelectronic package. Characterizing the fracture behavior of materials at the micrometer scale is becoming more relevant and important due to the continuing miniaturization of electronic devices[14]. Utilizing *in-situ* SEM techniques open up a number of possibilities to track the deformation and fracture behavior of a specimen at a scale that would otherwise not be visible using a light optical microscope or would be limited to characterizing after a macro scale test[15]. This technique aids in the understanding of fundamental deformation mechanisms that are critical to the brittle behavior of materials that exhibit a ductile to brittle transition.

The primary advantage of the edge-notched clamped beam geometry is that it allows observation of the crack trajectory across the beam as well as how the crack interacts with second phase particles and grain boundaries within the Sn matrix. In this manuscript we describe the use of this technique using a custom *In-situ* cryogenic mechanical testing apparatus, coupled with EBSD for characterizing the mechanical behavior below the DBTT.

## 2. Results and Discussion

### 2.1 *Cryogenic Mechanical Testing*

A major benefit to using the clamped beam geometry is that it is possible to characterize the deformation mechanisms of the Sn96 ahead of a preexisting flaw or in this case a crack. As discussed in more detail in the Methods section, this study focused on four different and unique microstructural configurations. Each configuration was chosen in order to characterize the interaction between the crack and any intermetallics, the Sn matrix and cracks initiated within an intermetallic matrix.

It has been previously shown that the increase of Ag content shifts the ductile to brittle transformation temperature towards warmer temperature, from -125C in the case of pure Sn to -45C for Sn96[7]. Ag has low solubility in Sn and instead will precipitate as $SnAg_3$ particles, the volume fraction of these particles will increase with increasing Ag content. The presence of these particles in the solder matrix have been considered to act as obstacles for dislocation motion as well as serve as sources for crack nucleation[7]. $SnAg_3$

particles can therefore act as obstacles to slip and have a very significant effect on the fracture behavior of the solder. For this reason, this study placed more focus on the effect of the large intermetallic particles in the matrix.

Figure 1 summarizes the mechanical response of sample 1, which did not contain any intermetallic particles ahead of the crack tip. Sample 1 was loaded twice as during the first load-controlled test there was no significant opening of the crack tip, deformation was primary accommodated along the outer edges of the beam. Whereas after the second load test (Figure 1b) there was considerable opening of the crack tip. Figure 1d shows the position of the diamond wedge tip with respect to the sample and Figure 1a shows the resulting deformation in sample 1.

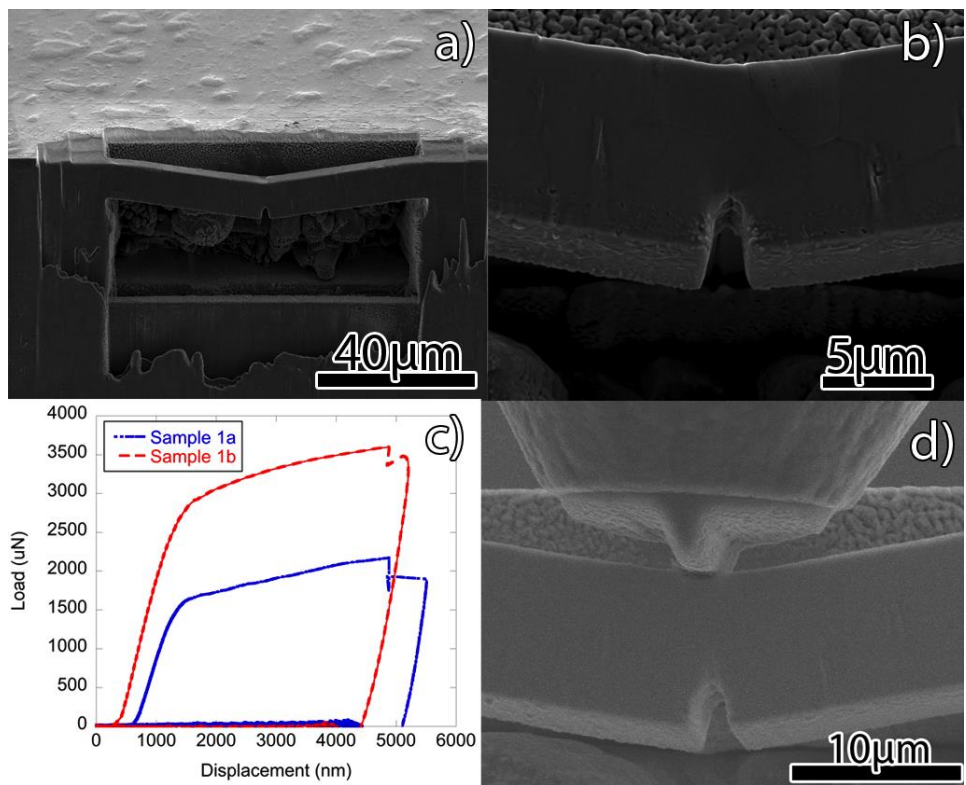

Figure 1: (a) SEM image of sample 1 after completing the bend test, (b) higher resolution image of the crack in sample, (c) the load and displacement curves for sample 1 after the first (1a) and second (1b) load cycle. (d) The position of the diamond wedge with respect to sample 1 during bending.

The mechanical response of sample 2 is shown in Figure 2. Sample 2 contained an intermetallic particle ahead of the crack tip. The intermetallic particle was both larger then the diamond wedge tip and in direct contact with the tip upon loading. Figure 2 shows the resulting deformation of sample 2. In addition to the crack tip opening upon loading, slip traces can also be seen within the intermetallic particle (Figure 2b) and will be shown in more detail in the following section.

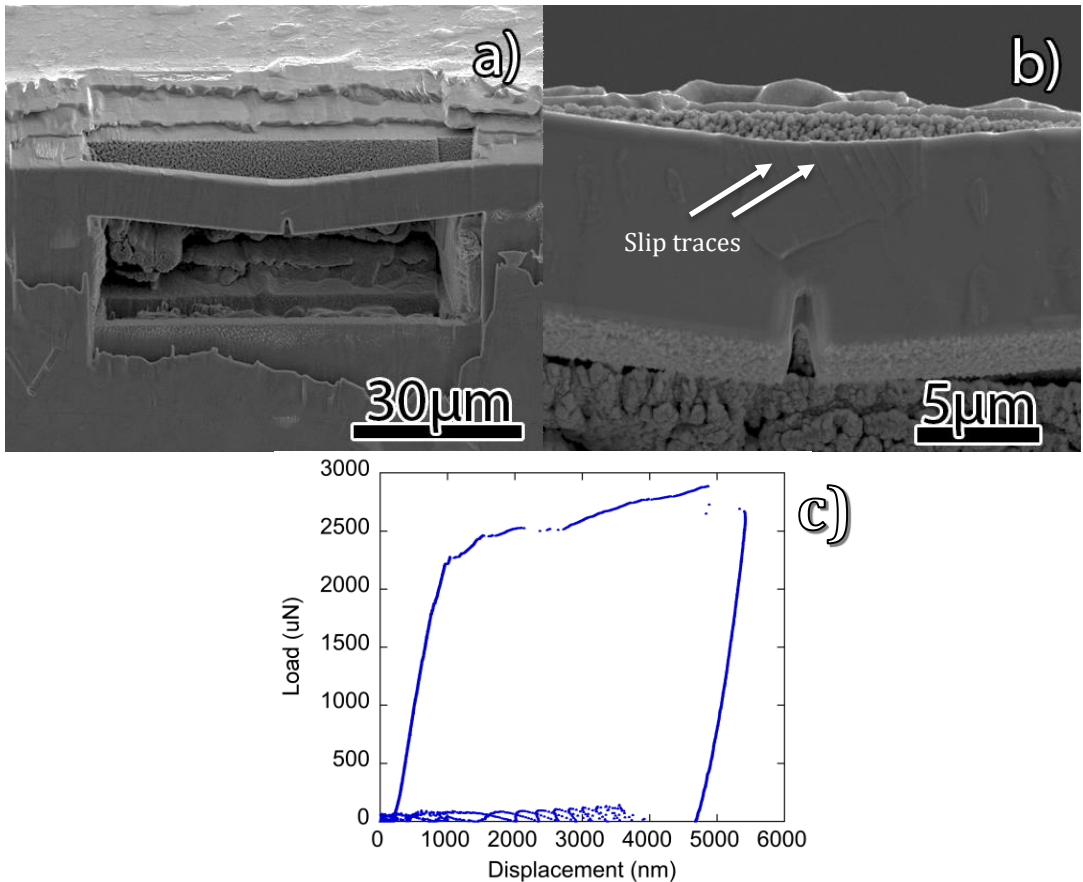

Figure 2: SEM image of sample 2 after completing the bend test (a), higher resolution image of the crack in sample 2 (b), and the load and displacement curves for sample 2 (c).

The mechanical response of sample 3 is shown in Figure 3. Sample 3 is similar to sample 2 with the exception that the intermetallic particle in sample 3 is smaller than the particle in sample 2. Figure 3b shows the resulting deformation of sample 3, which shows the crack running directly into the intermetallic particle and then continuing along the intermetallic boundary. Compared to the load/displacement curve of sample 2, sample 3 appears to only accommodate deformation through the opening of the crack tip, as the load significantly drops once the crack tip opens.

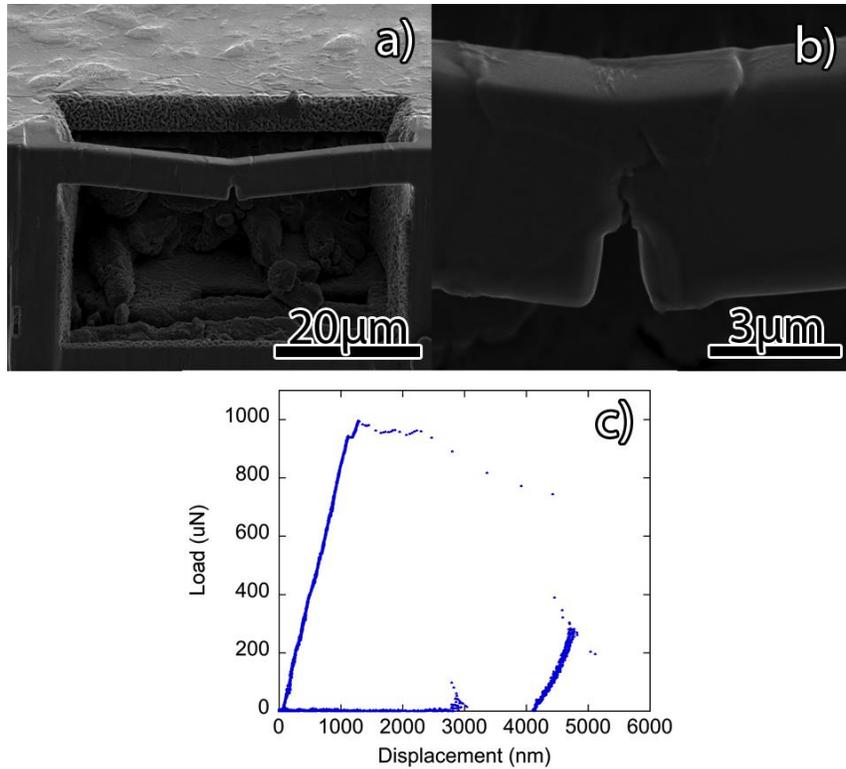

Figure 3: SEM image of sample 3 after completing the bend test (a), higher resolution image of the crack in sample 3 (b), and the load and displacement curves for sample 3 (c).

The mechanical response of sample 4 is shown in Figure 4. Unlike the previous samples where the crack was initiated within the Sn matrix, in this case the crack tip is initiated within the intermetallic phase. Directly above the intermetallic particle where the crack was initiated, was another intermetallic particle as shown in Figure 4a and 4b (annotated with arrows in Figure 4b). Upon deformation and crack tip opening, the crack progressed along the boundary between the two intermetallic particles rather then through the particle. This suggests that in the presence of the intermetallic phase, intergranular fracture is preferred and can contribute to the brittle behavior observed in Sn96 below the DBTT.

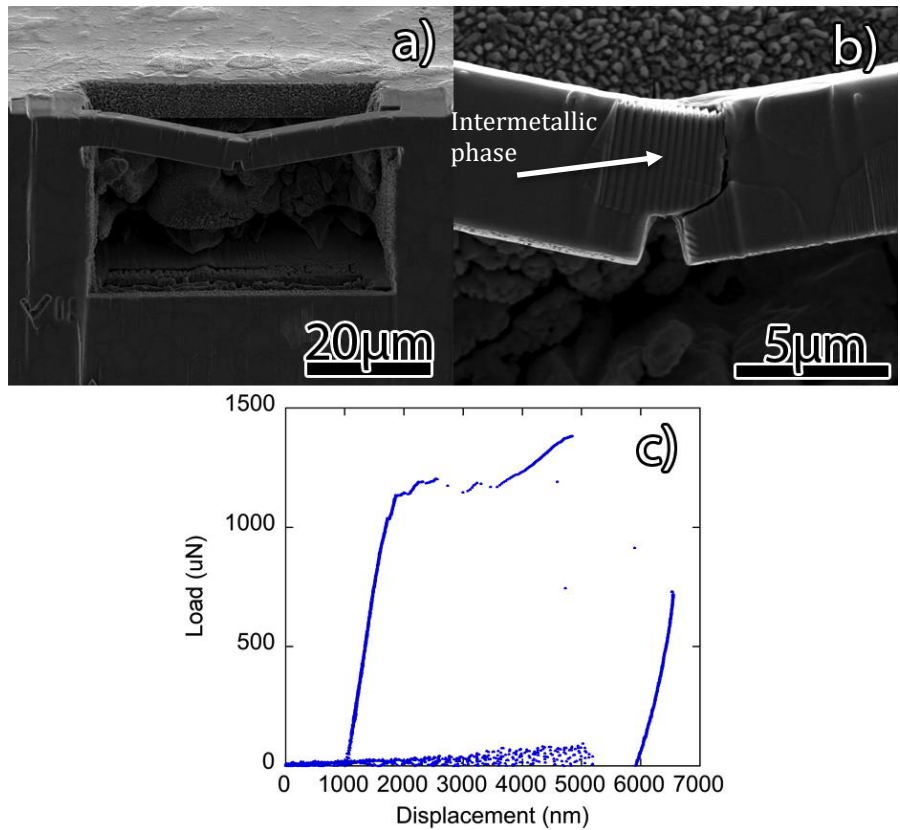

Figure 4: SEM image of sample 4 after completing the bend test (a), higher resolution image of the crack in sample 4 (b), and the load and displacement curves for sample 4 (c).

3.2 *EBSD of the Clamped Beam Geometry Specimens*

All of the microbeams were examined with EBSD prior to and directly after mechanical testing. Figure 5 shows the EBSD scans for sample 1 where part (a) shows the phase map where red represents the Sn matrix and blue represents the $Ag_3Sn$ intermetallic, part (b) shows the IPF Y map prior to testing and part (c) shows the IPF Y map directly after testing and part (d) shows an IPF Y map of the crack tip after loading. Prior to bending, sample 1 did not show any evidence of twinning, however as can be seen in Figure 5c, after bending below the DBTT, twins are present both ahead of the crack tip and at the edges of the beam. Twin boundaries are highlighted in white in the EBSD map. This is in agreement with twins forming in regions of highest stress concentration within the beam. Figure 5d shows the twin boundaries ahead of the crack tip in more detail. Similar to the twins previously observed in the pure Sn micropillars[1], twinning behavior is consistent with the formation of the {301} and {101} twin in Sn[17,18].

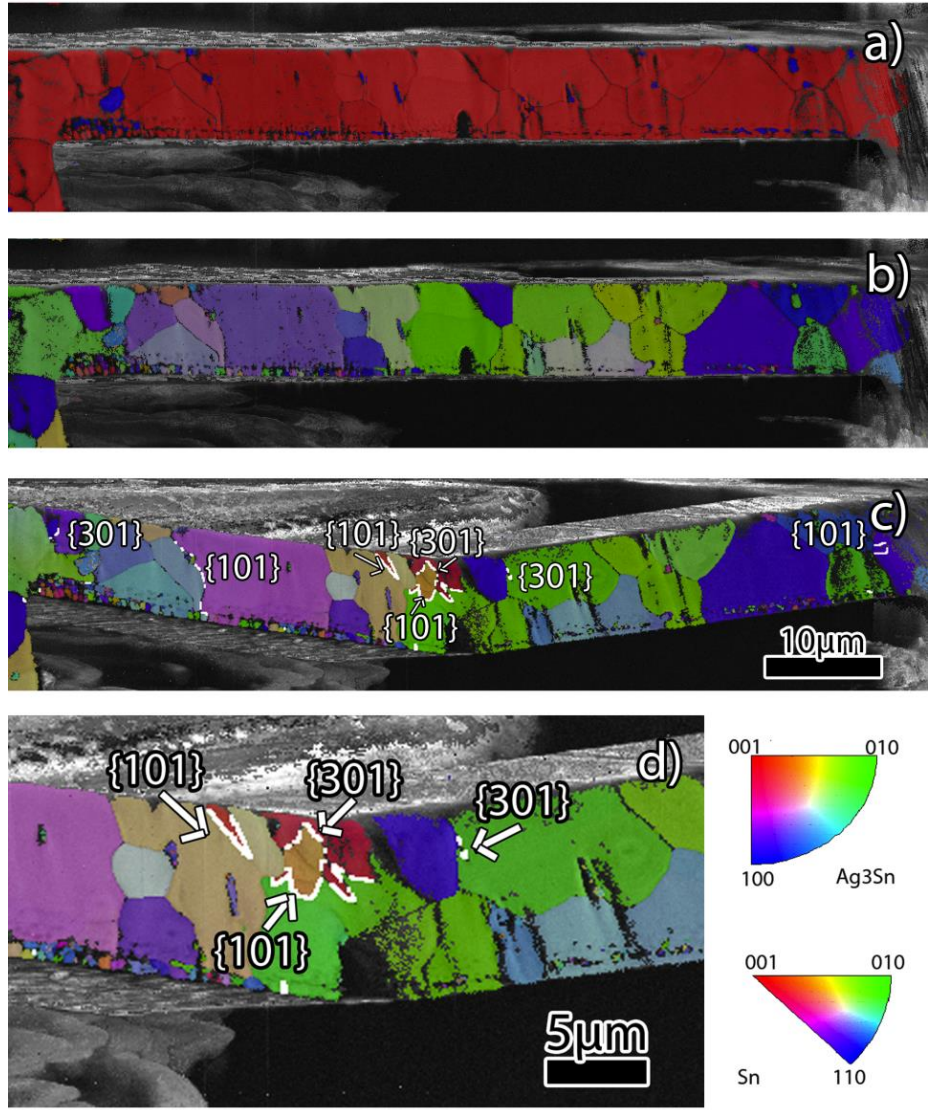

Figure 5: Clamped beam geometry sample 1 (a) phase map, (b) IPF Y map prior to bending, (c) IPF Y post bending and (d) IPF Y map of the crack tip post bending.

The EBSD scans for sample 2 are shown in Figure 6. As can be seen in Figure 6c twins also formed ahead of the crack tip and along the edges of the beam at the points of highest stress concentration. Figure 6d shows the deformation ahead of the crack tip in more detail where both {101} and {301} twins are observed. This suggests that the primary deformation mechanism ahead of the crack tip is deformation twinning and the required stress for the nucleation of dislocations at the crack tip is larger than the stress required to nucleate twins. Deformation is also accommodated within the intermetallic phase itself. Figure 6d also shows slip traces within the intermetallic phase that were not present prior to bending. While dislocations are not favorable in the BCT Sn matrix, it appears that they are still able to form and accommodate slip within the orthorhombic intermetallic phase.

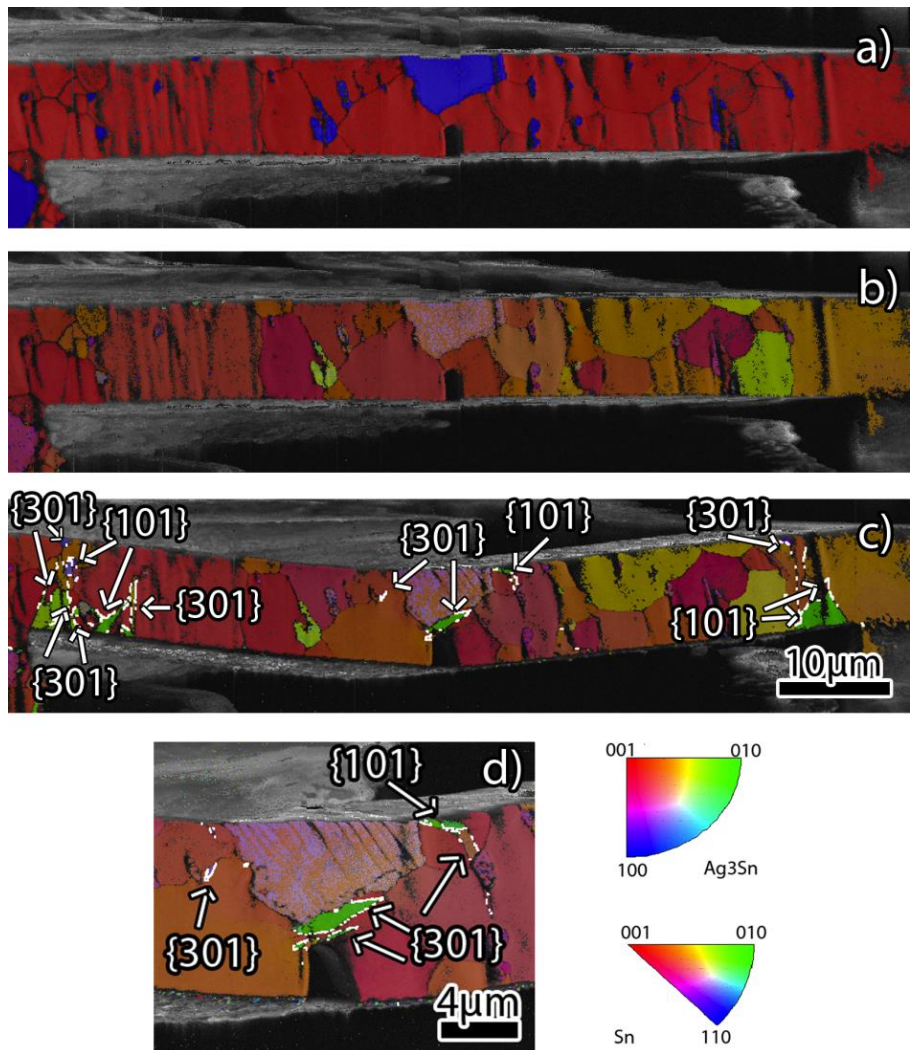

Figure 6: Clamped beam geometry sample 2 (a) phase map, (b) IPF X map prior to bending, (c) IPF X post bending and (d) IPF X map of the crack tip post bending.

The EBSD scans for sample 3 are shown in Figure 7. Sample 3 is similar to sample 2 in that the crack tip is initiated within the Sn matrix and ahead of the intermetallic phase. However unlike sample 2, sample 3 does not show any twinning ahead of the crack tip. Instead, deformation is accommodated through a small grain rotation as can be seen in Figure 7c and d. In this case the crack propagates initially along the second phase boundary and then stops at a slip step in the intermetallic phase. Twinning is observed in sample 3 along the beam edges, both being {301} twins as shown in Figure 7c.

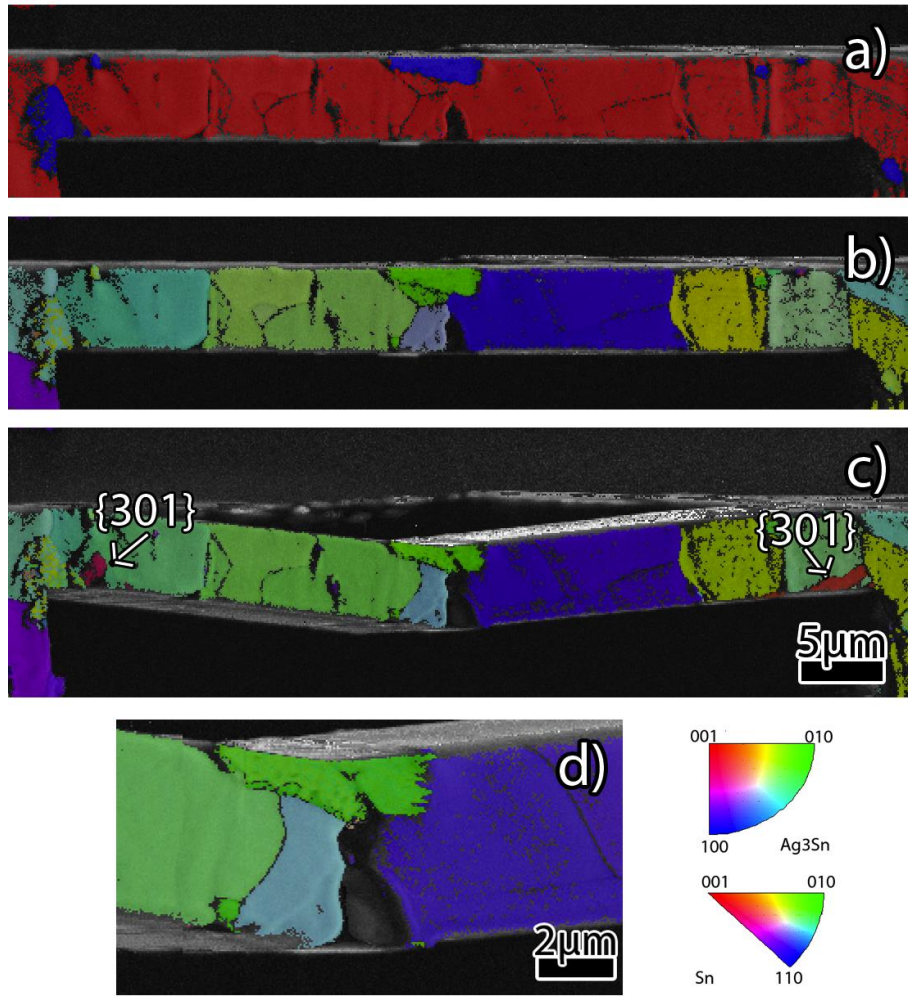

Figure 7: Clamped beam geometry sample 3 (a) phase map, (b) IPF Y map prior to bending, (c) IPF Y post bending and (d) IPF Y map of the crack tip post bending.

The EBSD scans of sample 4 are shown in Figure 8. Figure 8a clearly shows where the crack was initiated within the intermetallic phase, which is ahead of a second intermetallic particle of a different orientation (Figure 8b). Upon bending, the crack propagated to the second intermetallic particle boundary and continued to progress along the boundary between the two particles (Figure 8 c and d). This suggests that intergranular fracture, in particular between second phase particles, contributes to the brittle behavior of the Sn96 alloy below the DBTT. Twins were also observed in the neighboring Sn matrix grains that were directly next to the intermetallic particles as well as at the edges of the beam, which was observed in all four samples. The occurrence of twins in the surrounding Sn matrix suggests that the limited dislocation plasticity and deformation primarily through twinning is also a contributing factor to the brittle nature of the alloy. The twins observed, both the {301} and {101} are consistent with twins observed in pure Sn.

Using this clamped beam bending technique has made it possible to visualize and characterize toughening mechanisms near and around the crack.[11] Due to the stability of this geometry the crack trajectory was easy to visualize insight into the active deformation mechanisms was directly obtainable. While traditional bulk DBT and toughness testing methods are important in understanding global mechanical trends in the material, one advantage of micromechanical testing is the ability to target specific areas of interest that may affect the crack trajectory and deformation mechanisms near and around it as it progresses. In this study we were able to clearly see how cracks are directly affected by features that are inherent to the material such as second phase intermetallics.

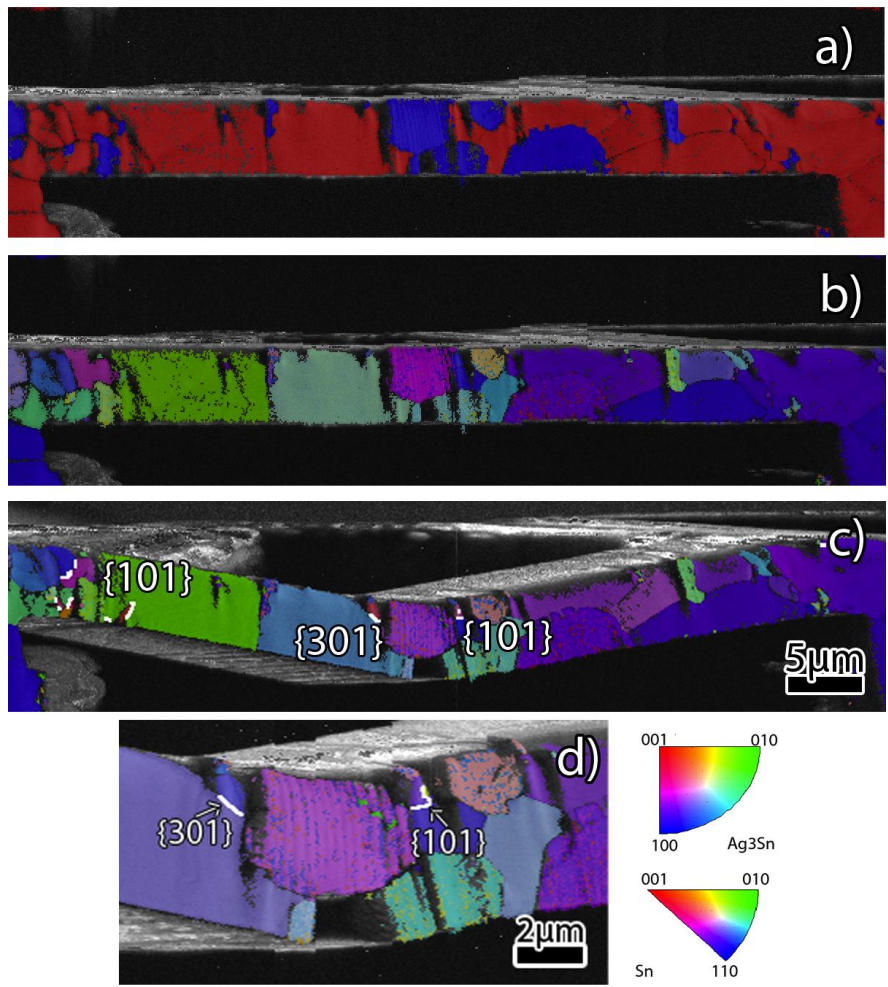

Figure 8: Clamped beam geometry sample 4 (a) phase map, (b) IPF Y map prior to bending, (c) IPF Y post bending and (d) IPF Y map of the crack tip post bending.

3. **Summary**

This work has demonstrated a new method for characterizing the mechanical behavior of alloys at cryogenic temperatures (-142°C). Furthermore this paper also demonstrated the use of EBSD both prior to and post deformation, allowing us to link the associated deformation mechanisms with behavior around and near an existing flaw, in this case the crack tip. Within the Sn matrix, as expected, deformation twinning is the primary deformation mechanism below the DBTT, occurring in the areas of highest stress concentration. However, deformation is not only accommodated within the Sn matrix. Deformation is also accommodated within the intermetallic phase itself where slip steps were observed within the intermetallic phase that were not present prior to bending.

**Availability of data and materials**

The datasets supporting the conclusions of this article are included within the article and available upon request from the authors.

## 4. Experiments and Methods

*In-situ* testing in a relevant environment allows us to link the deformation mechanisms of the material with the fracture behavior[11]. The geometry used in this study is a loaded edge-notched doubly clamped beam, which allows us to investigate the crack trajectory across the beam thickness. The methodology and fabrication of these specimens was adapted from previous studies done by Jaya et al.[11,13,14,15] The edge-notched clamped beam bend test geometry provides stability even under load control. This geometry was developed into a fracture test geometry as a modified form of three point bending.

This geometry facilitates direct resolution of the active deformation mechanisms near regions of interest, such as ahead of a grain boundary, ahead of an intermetallic phase particle, and within an intermetallic phase particle. In this study, edge-notched clamped beam geometry was applied to the Sn96 alloy, which allowed for direct characterization of how intermetallic phase particles affect crack behavior.

### 4.1 *Specimen Fabrication*

Before preparing the clamped beam specimen fabrication, the Sn96 sample was mechanically-polished on two adjacent faces. During polishing the Sn96 was mounted directly next to a thin piece of steel in order to prevent deformation and rounding near the edge of the Sn96. The samples were planarized using SiC grinding paper with water as a lubricant and were then polished with $0.3\mu$m, $0.1\mu$m alumina polishing solutions, and $0.05\mu$m colloidal silica polishing solution. This technique was also used and demonstrated in our previous studies on pure Sn[1].

Prior to ion beam machining, EBSD was used to identify grains for specimen locations. Fabrication of the doubly clamped beams was done using a FEI Quanta dual beam Focused Ion Beam/Scanning Electron Microscope (FIB/SEM) with a Ga$^+$ ion source operated at 30keV. The beams were fabricated close to the polished edge using an ion current of 1nA for coarse milling and 10pA for a final polish of the beam faces. A final polish was done on the top surface of the beam by tilting the sample to 90° in order to ensure that top of the beam is flat and uniform. The clamped beam is essentially a bridge specimen, where the region of interest is elastically fixed at the two ends to its parent material. A notch is inserted at the bottom center and loaded in bending, up to fracture[16]. The length (L) to width (W) ratio in all of these beams was maintained to be greater than 4 while crack length (a) to width (W) ratio was between 0.3 and 0.4. Beam dimensions were chosen on the basis of both the microstructural feature being tested (i.e an intermetallic) and small-scale yielding criteria. Four different beams were evaluated, each with a unique microstructural configuration. The dimensions of each beam as well as a description of each specimen are summarized in Table 1 below. The purpose of each configuration was to characterize the interaction between the crack and any intermetallics, the Sn matrix, and cracks intiated within an intermetallic material.

| Specimen | Length (L) μm | Width (W) μm | Thickness (t) μm | Crack Length (a) μm | Description |
| --- | --- | --- | --- | --- | --- |
| 1 | 84.1 | 9.32 | 4.80 | 2.80 | Crack was not ahead of any intermetallics |
| 2 | 72.4 | 9.24 | 3.81 | 3.15 | Crack was ahead of Ag3Sn intermetallic |
| 3 | 49.6 | 5.19 | 2.69 | 1.83 | Crack was ahead of Ag3Sn intermetallic |
| 4 | 55.2 | 5.34 | 3.72 | 1.67 | Crack initiated within Ag3Sn intermetallic |

Table 1. Summary of clamped beam specimens and their corresponding geometry and description.

**4.2** *Cryogenic Micromechanical Testing*

The clamped beam specimen tests were performed *In-situ* in the FIB using a Hysitron PI-85 nanoindenter. The FIB machined clamped micro-beams were loaded using a wedge tipped nanoindenter. The indenter was blunt enough to not produce permanent impressions on the specimens, so as to generate pure bending[8,9]. The radius of the wedge was approximately 1.5 $\mu m$. All of the clamped beam specimens were tested at -142°C (±2.5 °C) using a novel *In-situ* cryogenic stage that was previously demonstrated by Lupinacci et al[1]. To perform cryogenic testing, a custom-built cryogenic cooling system manufactured by Hummingbird Scientific was used in conjunction with the Hysitron PI-85 nanoindenter. The cryo system cooled the tip and sample simultaneously in order to minimize thermal drift. The validation of this system is demonstrated in our previous paper by Lupinacci et al[1]. It is necessary to perform these tests under vacuum in order to minimize and prevent the formation of ice on the specimen that can be caused by air humidity. Due to the soft nature of Sn96 at room temperature, this geometry was not useful above the DBTT since indentation rather than bending dominated the mechanical response. All testing was performed in load-control mode and the loading rate was kept constant at 0.5 mN/s. The load-displacement data and the real-time video of the bending deformation were captured and used for post experimental analysis.

*2.3 EBSD of the Clamped Beam Geometry Specimens*

EBSD was performed at room temperature to identify grains for clamped beam fabrication. After the specimens were mechanically-tested at -142°C, room temperature EBSD was used again to analyze the orientation of the clamped beam geometry specimens after deformation. Analysis was performed using Oxford/HKL data collection software at an accelerating voltage of 30 kV. A step size of 50nm was used for each sample. The collected data were used to construct inverse pole figure (IPF) maps. A low level of data interpolation was used with a criterion of 7 neighbors for extrapolation. Wild spikes and zero solutions were also removed. All maps were compared with band contrast maps to verify that the noise reduction did not lead to any spurious results.

## 5. Acknowledgments

Part of this research was carried out at the Jet Propulsion Laboratory, California Institute of Technology, under a contract with the National Aeronautics and Space Administration. AL was supported by a NASA GRP Fellowship and also by Boeing, Inc. We would like to thank both Hummingbird Scientific, Inc. and Hysitron, Inc. for help with the design and fabrication of the cryogenic testing apparatus. The authors acknowledge support from the Molecular Foundry at Lawrence Berkeley National Laboratory, which is supported by the U.S. Department of Energy under Contract No. DE-AC02-05-CH11231.